\documentclass[twocolumn]{openjournal}

\usepackage{newtxtext,newtxmath}
\usepackage[T1]{fontenc}

\makeatletter
\def\fnum@table{{\@eapj@cap@font Table~\thetable.}}
\long\def\@makecaption#1#2{%
 \noindent\begin{minipage}{0.9999\linewidth}
   \if\csname ftype@\@captype\endcsname 2 
   \vskip 2ex\noindent \centering\footnotesize{#1}~#2\par\medskip
   \else
   \vspace*{\abovecaptionskip}\noindent\footnotesize #1 #2\par\vskip \belowcaptionskip
   \fi
 \end{minipage}\par
}
\makeatother

\usepackage{graphicx} % Required for inserting images
\usepackage{amsmath}
\usepackage{listings}
\usepackage[dvipsnames]{xcolor}
\usepackage{hyperref}
\usepackage{orcidlink}
\usepackage{soul}
\usepackage{lineno}
\usepackage{needspace}
    [2021/06/11 v1.0.4 Linked ORCiD logo macro package]

\begin{document}
\title{Optimal intrinsic alignment estimators in the presence of redshift-space distortions}
\author{Claire Lamman\orcidlink{0000-0002-6731-9329},$^{1,2}$ Jonathan Blazek\orcidlink{0000-0002-4687-4657},$^3$ Daniel J. Eisenstein$^{1}$}
\affiliation{$^1$ Center for Astrophysics $|$ Harvard \& Smithsonian, 60 Garden Street, Cambridge, MA 02138, USA}
\affiliation{$^2$ Center for Cosmology and AstroParticle Physics (CCAPP), Ohio State University, Columbus, OH 43210}
\affiliation{$^3$ Department of Physics, Northeastern University, Boston, MA, 02115, USA}

\begin{abstract}
We present estimators for quantifying intrinsic alignments in large spectroscopic surveys that efficiently capture line-of-sight (LOS) information while being relatively insensitive to redshift-space distortions (RSD). We demonstrate that changing the LOS integration range, $\Pi_{\rm max}$, as a function of transverse separation outperforms the conventional choice of a single $\Pi_{\rm max}$ value. This is further improved by replacing the flat $\Pi_{\rm max}$ cut with a LOS weighting based on shape projection and RSD. Although these estimators incorporate additional LOS information, they are projected correlations that exhibit signal-to-noise ratios comparable to 3D correlation functions, such as the IA quadrupole. Using simulations from A\textsc{bacus}S\textsc{ummit}, we evaluate these estimators and provide recommended $\Pi_{\rm max}$ values and weights for projected separations of $1-100 h^{-1}$Mpc. These will improve measurements of intrinsic alignments in large cosmological surveys and the constraints they provide for both weak lensing and direct cosmological applications.
%Key: \todo{To Do} \cml{Claire's notes / questions}
\end{abstract}

\section{Introduction}

Large-scale structure is subtly imprinted in the orientations of galaxies. The most common form of this intrinsic alignment (IA) is for the long axis of elliptical galaxies to be aligned along the large-scale gravitational tidal field. Characterizing these correlations is critical for weak lensing studies; IA can bias the matter power spectrum by 30\% as measured by cosmic shear \citep{hirata_galaxy-galaxy_2004}, and cosmological results from photometric surveys are particularly sensitive to IA modeling \citep{kirk_galaxy_2015, dark_energy_survey_and_kilo-degree_survey_collaboration_y3_2023}. In principle, IA can also measure any cosmological effect that leaves an imprint on the large-scale tidal field, including primordial physics and the nature of dark energy \citep{chisari_cosmological_2013}. See \cite{joachimi_galaxy_2015, troxel_intrinsic_2015, chisari_rising_2025} for reviews and \cite{lamman_ia_2024} for a practical guide to IA estimators and formalisms. Spectroscopic surveys, combined with imaging data, provide the best direct measurements of IA, including the Sloan Digital Sky Survey (SDSS) \citep{singh_intrinsic_2015} and first year of data from the Dark Energy Spectroscopic Instrument (DESI) \citep{siegel_intrinsic_2025}. However, common conventions for measuring IA were largely developed with lensing in mind \citep{mandelbaum_detection_2006}. The advent of large spectroscopic surveys like DESI, with their potential to yield insights on IA for both upcoming imaging surveys and direct applications, means it is time to revisit these estimators.

IA are most commonly measured as projected statistics. The projected shapes of galaxies are measured relative to the underlying matter density, as traced by galaxies, and these correlations are averaged over a total LOS distance $2\Pi_{\rm max}$. Common $\Pi_{\rm max}$ choices are 60 - 100 $h^{-1}$Mpc. Some advantages of these projected statistics are that they are less sensitive to LOS uncertainties and can be more directly related to weak lensing measurements, which are sensitive to transverse modes. As with galaxy clustering, several recent studies explore higher-order statics which take advantage of the full 3D information provided in spectroscopic surveys \citep{schmitz_time_2018, kurita_power_2021, pyne_three-point_2022, bakx_bispectrum_2025}. Unlike galaxy clustering, the nature of tidal alignments and projected shapes leads to most information lying along the transverse direction (Figure \ref{fig:diagram}). In the case of tidal alignment, on average the long axis of galaxies are oriented towards a tracer. Because we observe only the projected shape of galaxies, this results in a weaker signal along the LOS, especially at small transverse separations. Therefore, an IA quadrupole estimator that essentially up-weights the signal in the transverse direction can increase the signal-to-noise ratio (SNR) of measurements \citep{singh_increasing_2024}. 

A disadvantage of using these 3D estimators in practice is any uncertainty in the LOS position of galaxies, primarily redshift-space-distortions (RSD). In redshift space, galaxy peculiar velocities create a ``smearing'' along the LOS below separations of about 3 $h^{-1}$Mpc \citep{jackson_critique_1972}, known as the fingers-of-god effect (FOG). On larger scales, infall into overdense regions produces a  LOS ``squashing'' (Kaiser 1987), known as the Kaiser effect. While this does not affect intrinsic shear to linear order \citep{singh_intrinsic_2015}, it does impact  nonlinear features which are integral to many direct cosmological applications of IA \citep{chen_lagrangian_2024, matsubara_integrated_2024, okumura_nonlinear_2024}. Since the IA signal does not follow a perfect $\mu$ relation in redshift space, 3D IA correlations must successfully model these effects.

Here we present a set of alternative estimators that preserves the main advantages of both classic and 3D measurements: a $\Pi_{\rm max}$ which varies with transverse separation and a LOS weighting based on both shape projection and RSD. They have the potential to provide similar SNR as the IA quadrupole while being less sensitive to redshift errors and relatively insensitive to RSD. 

To develop and evaluate these new estimators, we use halo catalogs from the A\textsc{bacus}S\textsc{ummit} simulations \citep{maksimova_abacussummit_2021}, a suite of large, high-accuracy simulations designed to meet the analysis requirements of DESI. While halos display stronger alignment than galaxies, the alignment strength itself will not significantly impact the relative performance of these estimators. The strong alignment and large volume of these simulations is well-suited to making high-precision determinations of signals and relative errors.

Throughout the paper we assume the cosmological parameters of $H_0=69.6$, $\Omega_{m,0}=0.286$, $\Omega_{\Lambda,0}=0.714$.

\vspace{.1in}\section{IA estimators}
\label{sec:estimators} 

\subsection{Existing formalism}

\begin{figure}
\centering
\includegraphics[width=.45\textwidth]{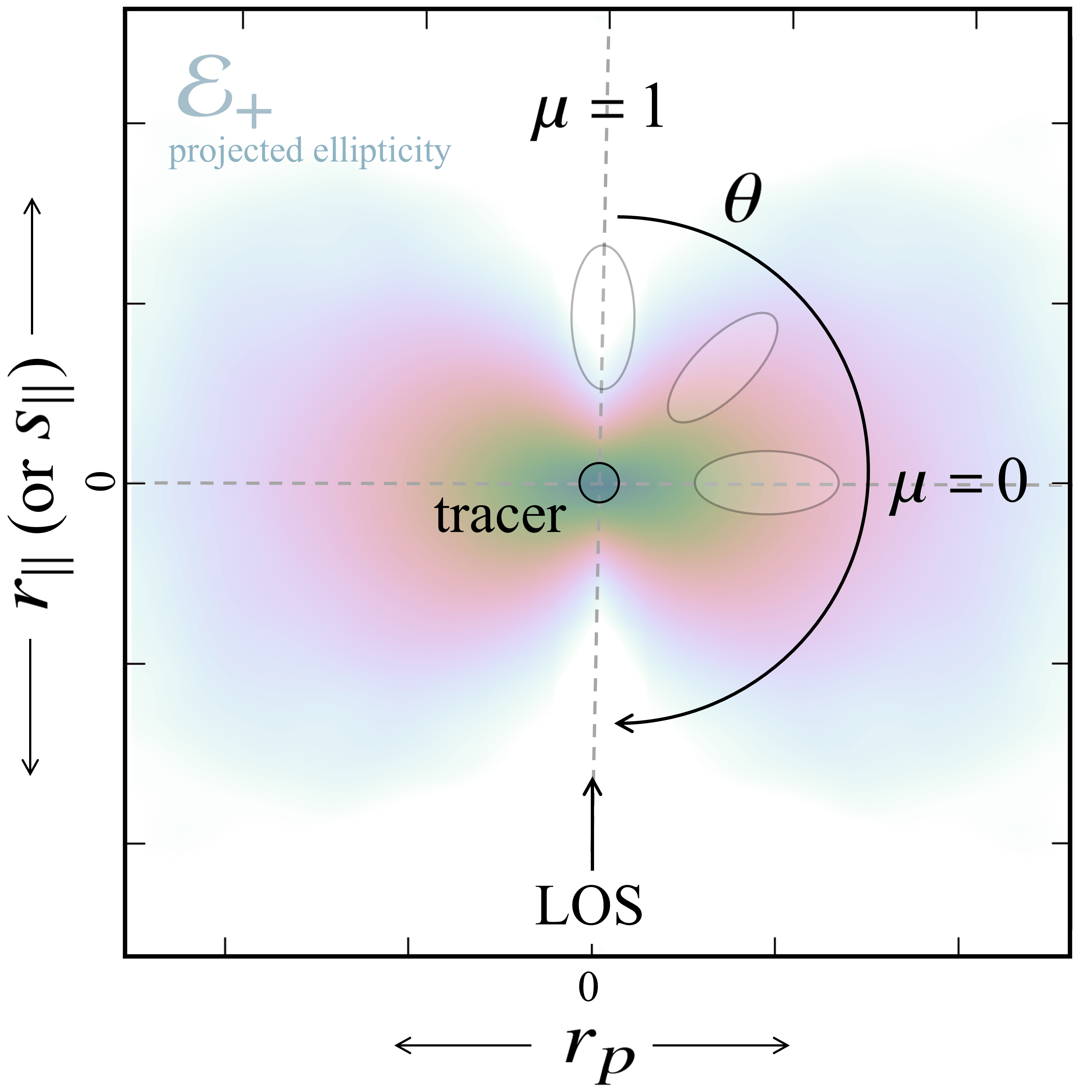}
\caption[]{A schematic of coordinates used in measuring IA. $r_p$ is the transverse distance relative to a central tracer, $r_\|$ is the distance along the LOS in real space, and $s_\|$ is the distance in redshift space. The background shading corresponds to the strength of measured tidal alignment; the extent to which a galaxy at that position will be alignment relative to the tracer. This is further illustrated by the outlines of three cartoon galaxies. Because we observe projected shapes, the measured alignment will be strongest along $r_p$ and approaches 0 along $r_\|$ at small $r_p$. The ellipticity measurements shown here are from the A\textsc{bacus}S\textsc{ummit} simulations, made in linear bins out to 100$h^{-1}$Mpc in $r_p$ and $r_\|$.\vspace{.07in}}
\label{fig:diagram}
\end{figure}

IA is quantified through a class of correlations involving galaxy positions and intrinsic shapes. Here we will exclusively focus on the intrinsic shape - density correlation, which describes the correlation between the projected shapes of galaxies and underlying density, which is typically traced by galaxy positions. Variables are summarized in Table \ref{tab:variables} and marked in Figure \ref{fig:diagram}. We use $r_\perp$ to describe transverse distance, $r_\|$ to describe LOS distance in real space, and $s_\|$ to describe LOS distance in redshift space. $r$ is the 3D separation and $\mu$ the angle relative to the LOS.

\begin{table}
\begin{center}
\begin{tabular}{|c|l|c|}
\hline
{\textbf{Variable}} & {\textbf{Description}} & Eq.\\ \hline
$r_p$ & projected separation, transverse to the LOS & \\
$r_\|$ & separation along the LOS in real space & \\
$s_\|$ & separation along the LOS in redshift-space & \\
$r$ & 3D real-space separation, $r^2 = r_p^2 + r_\|^2$ & \\ 
$\theta$ & angle with respect to the LOS, $0 < \theta < \pi$ & \\
$\mu$ & characterizes $\theta$, $\mu = \cos{\theta}$ & \\
\hline

$\mathcal{E}_+(r)$ & projected ellipticity relative to tracer. & \ref{eq:rele}\\
$\xi_{g+}$ & IA correlation function & \ref{eq:xi_gp}\\
$w_{g+}$ & projected IA correlation function & \ref{eq:w_g+}\\
$\tilde{w}_{g+}$, $\tilde{\mathcal{E}}_+$ & $s_\|$-weighted IA correlation functions  & \ref{eq:weighted_erel}\\
\hline
\end{tabular}\caption{A guide to variables used frequently throughout this paper.}
\label{tab:variables}
\end{center}
\end{table}

\subsubsection{Relative Ellipticity}
$\mathcal{E}_+(r)$ is the average projected ellitpicity relative to a neighbor galaxy as a function of separation $r$ \citep{lamman_intrinsic_2023}. It can be thought of as average intrinsic shear, $\epsilon_+$. This is averaged over $N$ shape-tracer pairs. For a shape catalog $S$ and tracer catalog $D$, 
\begin{equation}\label{eq:rele}
    \mathcal{E}_+(r) = \frac{1}{N}\sum\limits_{i \in S,j \in D} \epsilon_+(j|i) = \frac{S_+D(r)}{DD(r)},
\end{equation}
$DD$ and $S_+D$ are pair and weighted pair counts binned in $r$. This can also be averaged over bins of $r_p$ and/or $s_\|$ to obtain completely projected correlations. The relative ellipticity used to weight the counts in $S_+D$ is quantified as $|{\epsilon}|\cos(2\phi)$. $\phi$ is the on-sky angle between the galaxy orientation and separation vector to a tracer galaxy and $|\epsilon|$ is the absolute value of the shape's projected complex ellipticity. We define this in terms of the galaxy's projected axis ratio, $q$, as $( 1 - q) / (1 + q)$, but some other works use $( 1 - q^2) / (1 + q^2)$. Our results for optimal estimators do not depend on this choice. 

$\mathcal{E}_+$ is the estimator we use to determine weights and compare $\Pi_{\rm max}$ methods as it is related to the most relevant quantity: shape alignment, independent of clustering. Assuming IA noise is dominated by shapes and not clustering (as is almost always the case), the SNR which we use to compare estimators will be the same for both $\mathcal{E}_+$ and the more commonly used projected IA correlation function $w_{g+}$.

\subsubsection{IA correlation function}

The IA correlation function incorporates the 2-point galaxy clustering correlation function, $\xi(r)$ and uses random catalogs for the shape and tracer samples, $R_S$ and $R_D$ \citep{mandelbaum_detection_2006}. It is given as:

\begin{equation}\label{eq:xi_gp}
    \xi_{\rm g+} = \frac{S_+D - S_+R_D}{R_SR_D} \approx \frac{S_+D}{R_SR_D}
\end{equation}
This is a generalized form of the Landy-Szalay estimator \cite{landy_bias_1993}. The key difference compared to $\mathcal{E}_+$ is that $\xi_{\rm g+}$ is not averaged on a per-pair basis. They are related as
\begin{equation}
    \xi_{\rm g+}(r) = \frac{DD}{R_SR_D}\mathcal{E}_+(r) = \big(1+\xi(r)\big)\mathcal{E}_+(r).
\end{equation}

\noindent Especially for observations, this is often projected along the LOS to obtain the IA projected correlation function (e.g. \cite{blazek_tidal_2015}):
\begin{equation}\label{eq:w_g+}
w_{g+}(r_p) = \int_{-\Pi_{\rm max}}^{\Pi_{\rm max}}dr_\| \xi_{g+}(r_p, s_\|) \approx \big(2\Pi_{\rm max} + w_p(r_p)\big)\mathcal{E}_+(r_p).
\end{equation}
Here, $w_p$ is the 2-point projected clustering correlation function integrated along the same $\Pi_{\rm max}$ bounds. This can generally be related to the $\mathcal{E}_+$ estimator through $\big(2\Pi_{\rm max} + w_p(r_p)\big)$, but differences in binning pairs along the LOS can propagate as $O(0.1)$ amplitude shifts. $w_{g+}$ is the most common convention for direct IA measurements, and $\Pi_{\rm max}$ is typically 60 - 100 $h^{-1}$Mpc. While this estimator may be suitable for photometric surveys where there is a large LOS uncertainty, it is known to lose valuable LOS information when measuring IA in spectroscopic surveys.

\subsubsection{IA Multipoles}
Higher-order correlations, such as the IA quadrupole, seek to reclaim LOS information lost in projected statistics. Unlike galaxy clustering, the tidal alignment of galaxies is much stronger when $r_p > r_\|$ due to the geometry of projected shapes. This effect follows a $1-\mu^2$ dependence, as demonstrated in Figure \ref{fig:mu_dependence}, and is a result of projecting a 3D quadratic form onto a 2D plane. The projected shape orientation is not just a direct projection of the original axes, but a transformation of the entire ellipsoid surface, which involves a $\sin^2(\theta)=1-\mu^2$ term. 
\citet{singh_increasing_2024} proposes a ``weighted quadrupole'', which gives higher weight to shape-tracer pairs when $\mu$ is small. The signal is measured as a function of $r$, made by summing over bins of $\mu$ and weighting by $1-\mu^2 \propto L^{2,2}$:
    \begin{equation}
        \tilde{\xi}_{g+,2}(r) = \xi(r)\frac{\sum_{i\in\mu} (1 - \mu_i^2)\mathcal{E}_{+}(r, \mu_i)}{\sum_i^n(1-\mu_i^2)}
    \end{equation}
    
\noindent \cite{singh_increasing_2024} also removes all signal with $r_p<5h^{-1}$Mpc to avoid noise from FOG at low separations and the difficulty of modeling IA in the nonlinear regime. From Equation 18 of that work, the ``wedged'' quadrupole for $\xi_{g+}$ is:

\begin{equation}
\tilde{\xi}_{g+}^{2,2}(r) = \frac{5}{48} \int d\mu_{r} \Theta(r_p)L^{2,2}(\mu_{r})\xi_{g+}(r, \mu_{r})
\end{equation}

\begin{figure*}
\centering
\includegraphics[width=1\textwidth]{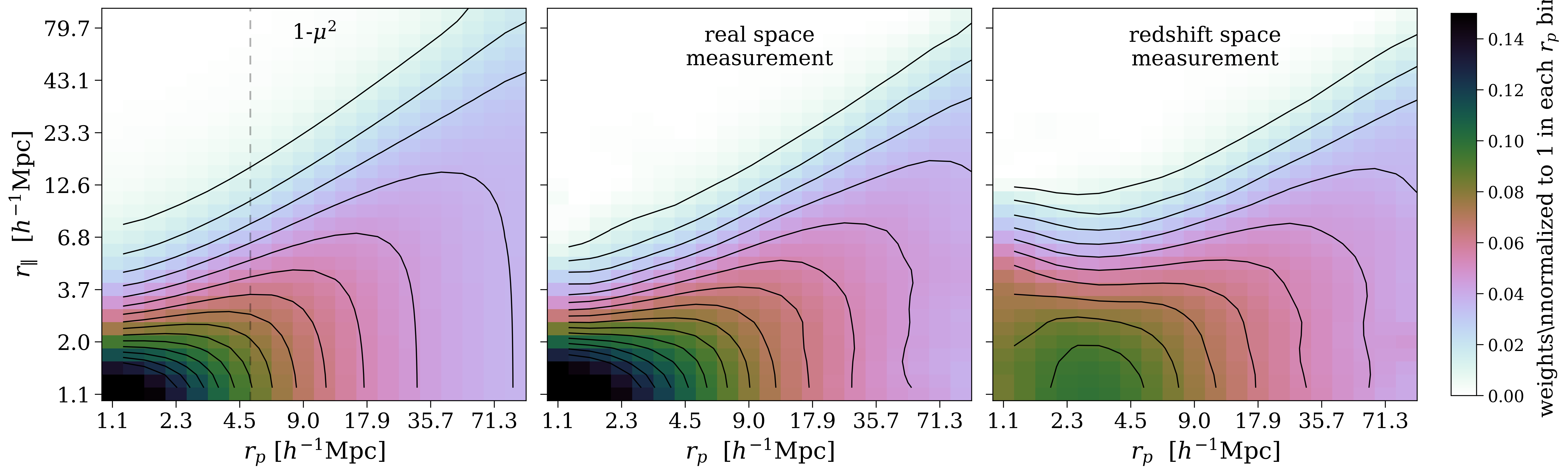}
\caption[]{Comparisons of tidal alignment and their angular dependence in real and redshift space. The first panel shows the mathematical relation $1-\mu^2$. This is a visualization of the weighted quadrupole used in \cite{singh_increasing_2024}, which is cut off below 5 $h^{-1}$Mpc. This is well-reproduced by the measurement of tidal alignment, $\mathcal{E}_+$ in real space (center panel), except for second-order effects from the $r_\|$ dependence IA. The right panel displays alignment binned in redshift space, which smears the signal at low separations. The signal in each panel is made in logarithmic bins of projected separation $r_p$ and either LOS distance in real space, $r_\|$, or redshift space $s_\|$. All signals are normalized in each $r_p$ bin to highlight the LOS difference across all scales.}
\label{fig:weight_derivations}
\end{figure*}

\begin{figure*}
\centering
\includegraphics[width=1\textwidth]{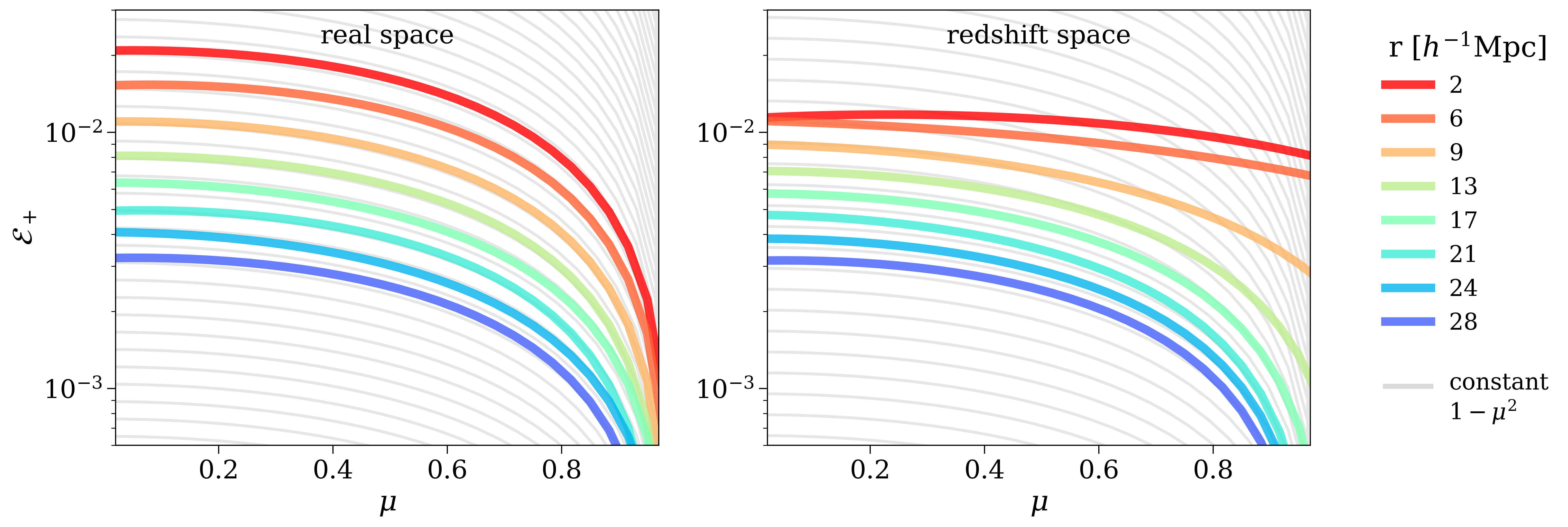}
\caption[]{An alternative visualization to Figure \ref{fig:weight_derivations}. Here, each colored line is the mean  $\mathcal{E}_+$ measured in a radial annulus with average distance from tracer $r$ as a function of $\mu$. Grey lines show contours of constant $1 - \mu^2$. Without RSD, the angular dependence of these signals is purely a result of shape projection, which follows $1-\mu^2$. With RSD, the signal is distorted, especially at close separations ($r<10$ Mpc/h) and close to the LOS ($\mu=1$). Each line is fit to measurements from A\textsc{bacut} S\textsc{ummit}. \vspace{.2in}}
\label{fig:mu_dependence}
\end{figure*}

\noindent Here, $L^{\ell,s_{ab}}$ are the Legendre polynomials of order $\ell=2$ and $s=2$. This is different from the quadrupole used to measure RSD, which uses $L_2 \propto \frac{1}{2}(3\mu^2 - 1)$. $\Theta(r_p)$ is the function that removes the contribution from separations below $5h^{-1}$Mpc. We demonstrate this weighting in Figure \ref {fig:weight_derivations}. $\mu$-based weightings will optimize SNR for triaxial ellipsoids with tidal alignment in real space. \cite{singh_increasing_2024} reports around 2 times greater precision using this wedged quadrupole compared to $w_{g+}$ with $\Pi_{\rm max} = 100h^{-1}$Mpc.

\subsection{Proposed estimators}
In the presence of redshift-space distortions, there are two additional factors we consider: the Kaiser effect, which smears positions along the LOS on large scales and the small-scale FOG regime below $r_p<5h^{-1}$Mpc. The Kaiser effect has a lower impact on the signal compared to FOG, since it distorts volumes in redshift space \citep{kaiser_clustering_1987,scoccimarro_redshift-space_2004}, leading to no linear effects in the intrinsic shear field \citep{singh_intrinsic_2015}. Although modeling is challenging in the FOG regime, most information can be found at small scales where IA is strongest. Figures \ref{fig:weight_derivations} and \ref{fig:mu_dependence} illustrate how the angular dependence of tidal alignment is affected by RSD using A\textsc{bacus} halos. The mathematical relation $1-\mu^2$ is replicated by tidal alignment in real space, but breaks down, especially at small separations, in redshift space. Here, close pairs of galaxies with strong alignments are smeared along the LOS. At very large $r$, the expected $1-\mu^2$ dependence is largely recovered. These plots emphasize the change in angular dependence, which is less pronounced in a large $r_p$ bin; the impacts of the Kaiser effect can be better seen in Figure \ref{fig:rsd_comp}. 

Here we explore two new ways to handle LOS information in each bin of $r_p$: a LOS weighting and a variable $\Pi_{\rm max}$. Both seek to mitigate the impact of RSD, preserve LOS information, minimize modeling challenges, and improve SNR of direct detections. Both also result in a familiar projected correlation function, similar to $w_{g+}$, and can be modeled similarly (Section \ref{sec:modeling}).

\vspace{.3in}
\subsubsection{LOS weighting}\label{sec:weighted_pimax}
A weighted $\Pi_{\rm max}$, or $\tilde{\Pi}$, functionally replaces the ``top hat'' of a set $\Pi_{\rm max}$ cut with a Gaussian. This Gaussian assigns a LOS weight to each shape-tracer pair within a given $r_p$ bin based on the signal dependence on shape projection and RSD. The result is a projected correlation function where each $r_p$ bin contains a unique LOS weighting. 

We define optimal weighting as one that maximizes the SNR of the final measurement and assume all noise comes from shape and shot noise. For a given $r_p$ bin, the quantity of interest is the SNR of the signal. This can be characterized by the alignment amplitude, or model re-scaling: $\mathcal{E} = \beta \mathcal{E}_{\rm model}$. The total signal summed over all pairs is proportional to $\mathcal{E}(1+\xi)$ and the noise is proportional to $\sqrt{1+\xi}$. The $\chi^2$ value for pairs in a given $r_p$ bin is summed over their transverse separations $s_\|$ as

\begin{equation}
    \chi^2_{r_p} = \frac{\big[\sum_{j\in s_\|}\mathcal{E}(1+\xi) - \sum_{j\in s_\|}\beta\mathcal{E}_{\rm model}(1+\xi)\big]^2}{\sum_{j\in s_\|}(1+\xi)}.
    %\chi^2_{r_p} = \frac{\big[\sum_{j\in s_\|}\mathcal{E}(r_p, j)\big(1+\xi(r_p, j)\big) - \sum_{j\in s_\|}\beta\mathcal{E}(r_p, j)_{\rm model}\big(1+\xi(r_p, j)\big)\big]^2}{\sum_{j\in s_\|}\big(1+\xi(r_p, j)\big)}.
\end{equation}
%\begin{equation}
%    \beta = \frac{\sum (SD) / \sigma^2}{\sum (DD) / \sigma^2} \rightarrow 
%    \tilde{\Pi}(r_\|) = \mathcal{E}(r_\|)
%\end{equation}
 $\mathcal{E}$ and $\xi$ are both functions of $r_p$ and the $j$ bin in $r_p$. To minimize this, we obtain weights $W$ which are proportional to the expected signal in redshift space $\mathcal{E}_+(r_p, s_\|)$. For a given $r_p$, the signal is a weighted sum across each pair, where the $s_\|$-dependent weight is determined individually for each $r_p$ bin:

\begin{equation}\label{eq:weighted_erel}
    \tilde{\mathcal{E}}_{+}(r_{p}) = \frac{\sum_{j\in s_\|}
    W(r_{p}, j)\mathcal{E}_+(r_{p}, j)}
    {\sum_{j\in s_\|}W(r_{p}, j)}.
\end{equation}

\begin{figure}
\centering
\includegraphics[width=.48\textwidth]{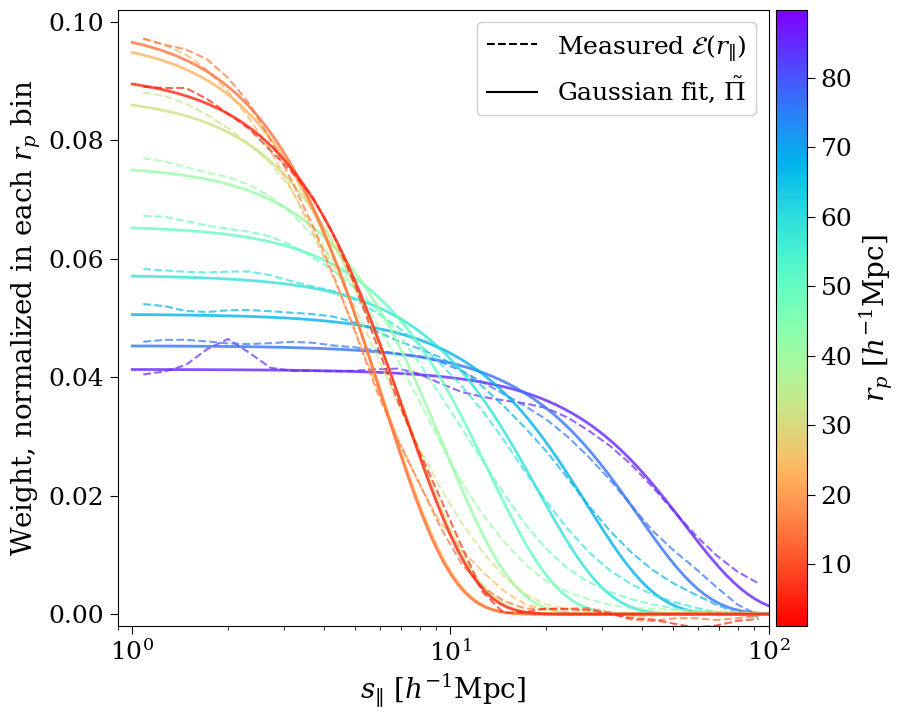}
\caption[]{A selection of the LOS weights as a function of $s_\|$ in each $r_p$ bin. The dashed lines are measured $\mathcal{E}_+$ and the solid are their Gaussian fits. At smaller projected separations, shown in red, higher weight is given to close pairs. At very small separations, below 3 $h^{-1}$Mpc, the functions slightly widen to account for FOG. This general widening of the functions at large separations is a result of the angular dependence of projected shapes, as displayed in Figure \ref{fig:weight_derivations}. The Gaussian widths are plotted in Figure \ref{fig:pimax_reccs}.}
\label{fig:weight_functions}
\end{figure}

It may seem unusual to weight a signal using the expected value of that signal, but the second-order scale-dependence of alignment will have a marginal impact on the determined weights; weights are normalized in each $r_p$ bin and the LOS-dependence in each $r_p$ bin is dominated by shape projection and RSD. Therefore we can also derive weights empirically using the A\textsc{bacus} halo catalogs (Section \ref{sec:abacus_tests}), which can be generalized to any elliptical galaxy population. Note that in real space, the derived weights will not quite recover the $1-\mu^2$ dependence along the $r_\|$ direction due to the scale dependence of the IA signal itself (this is subtly visualized in the difference between the first two panels of Figure \ref{fig:weight_derivations}). We determine $\mathcal{E}_+$ in 2D bins, shown in Figure \ref{fig:weight_derivations}, then fit a Gaussian to the $s_\|$-dependent signal to obtain weights for each $r_p$ bin. The measurements and fits are shown in Figure \ref{fig:weight_functions}. The Gaussians' widths are plotted in Figure \ref{fig:pimax_reccs} and can be found in Table \ref{tab:rp_wid1}. Assuming a linear relation, this follows $\sigma = 3 + (2/5)r_p$, with $R^2 = 0.99$ above projected separations of $r_p=2h^{-1}$Mpc. We explored fitting the curves with mixtures of 2-3 Gaussians but found no significant improvement in the final SNR. These weights can also be estimated analytically by combining the $1-\mu^2$ relation with models of RSD and $\mathcal{E}_+(r)$. %\todo{To Do: JB} \jab{Yes.}. 

\pagebreak
\subsubsection{Optimal $\Pi_{\rm max}$}\label{sec:optimal_pimax}
In cases where a traditional projected statistic is preferred, we recommend varying the $\Pi_{\rm max}$ for each $r_p$ bin. A version of this variable $\Pi_{\rm max}$ cut was used in \cite{lamman_detection_2024}, and can be measured and modeled with most existing IA infrastructure. We use the weights above to produce $\Pi_{\rm max}$ recommendations for given separation bins that maximize the SNR.

\begin{figure}
\centering
\includegraphics[width=.46\textwidth]{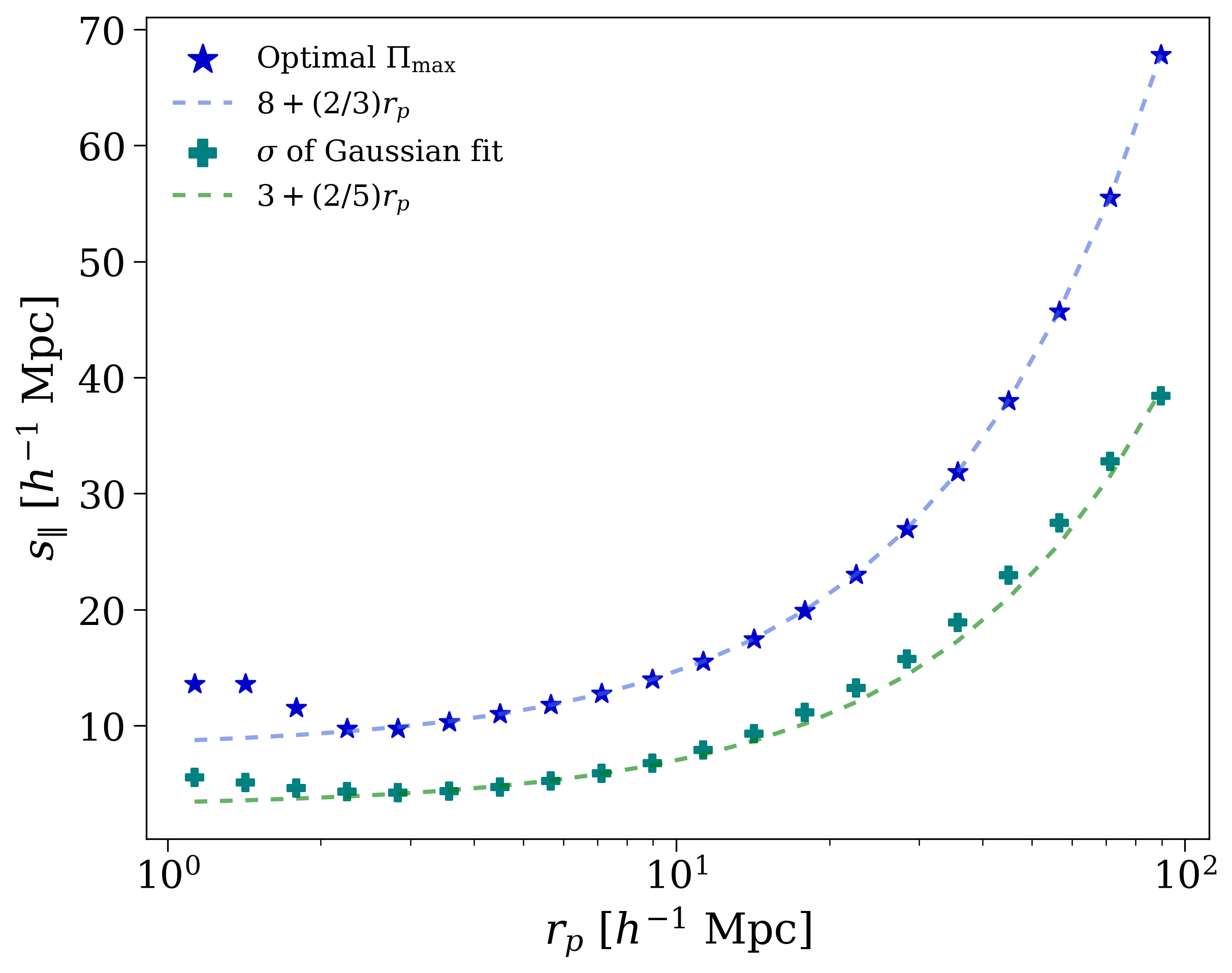}
\caption[]{A guide for how to treat the LOS direction in projected IA measurements. Generally, a larger $\Pi_{\rm max}$ choice is recommended for tidal alignment at larger projected separations, approximately following (2/3)$r_p$. Below 3-4 $h^{-1}$Mpc, a slightly larger $\Pi_{\rm max}$ is recommended as RSD smears the signal along the LOS. A Gaussian LOS-weight can further improve the signal. Widths of these Gaussians are shown as teal plus signs and based on the expected tidal alignment in a given $r_p$ bin (Figure \ref{fig:weight_derivations}). These approximately follow 3 + (2/5)$r_p$.}
\label{fig:pimax_reccs}
\end{figure}

For a slice in $r_p$, the total signal for a given separation $s_\|$ is the number of pairs times the average alignment: 

$\big(2s_\| + \xi(s_\|)\big)\mathcal{E}(s_\|)$. The shot noise is given by the \textbf{total} number of pairs within $s_\|$, $\sqrt{2\Pi_{\rm max} + w_p}$. For a given bin in $r_p$, the SNR 
as a function of $\Pi_{\rm max}$ is
\begin{equation}
    {\rm SNR}_{\Pi} = \frac{\big(2s_\| + \xi(s_\|)\big)\mathcal{E}}{\sqrt{2\Pi_{\rm max} + w_p}}.
\end{equation}
 The peak of this function, the $\Pi_{\rm max}$ choice to maximize signal-to-shot noise for IA measurements, is shown in Figure \ref{fig:pimax_reccs} and Table \ref{tab:rp_wid1}. This approximately follows the empirical relation $\Pi_{\rm max, optimal} = 8 + (2/3)r_p$, with $R^2 = 0.99$ above projected separations of $r_p=2h^{-1}$Mpc. It deviates in the FOG regime below 2-3$h^{-1}$Mpc. On these smaller scales, the optimal $\Pi_{\rm max}$ increases in order to capture information from close pairs which RSD have scattered along the LOS.

\begin{table}
\centering
\begin{tabular}{cc|cc}
    \hline
       $r_{p,\text{min}}$ & $r_{p,\text{max}}$ & $\sigma$ & Optimal $\Pi_{\rm max}$ \\
       \hline
    \multicolumn{4}{c}{[$h^{-1}$Mpc]} \\
    \hline
    1.0 & 1.3 & 5.5 & 13.6 \\
    1.3 & 1.6 & 5.1 & 13.6 \\
    1.6 & 2.0 & 4.6 & 11.5 \\
    2.0 & 2.5 & 4.3 & 9.7 \\
    2.5 & 3.2 & 4.2 & 9.8 \\
    3.2 & 4.0 & 4.4 & 10.3 \\
    4.0 & 5.0 & 4.7 & 11.0 \\
    5.0 & 6.3 & 5.2 & 11.8 \\
    6.3 & 7.9 & 5.9 & 12.8 \\
    7.9 & 10.0 & 6.8 & 14.0 \\
    10.0 & 12.6 & 7.9 & 15.5 \\
    12.6 & 15.8 & 9.3 & 17.5 \\
    15.8 & 20.0 & 11.1 & 19.9 \\
    20.0 & 25.1 & 13.3 & 23.0 \\
    25.1 & 31.6 & 15.8 & 26.9 \\
    31.6 & 39.8 & 18.9 & 31.9 \\
    39.8 & 50.1 & 23.0 & 38.0 \\
    50.1 & 63.1 & 27.5 & 45.7 \\
    63.1 & 79.4 & 32.8 & 55.5 \\
    79.4 & 100.0 & 38.4 & 67.8 \\
    \hline
\end{tabular}
\caption{The points plotted in Figure \ref{fig:pimax_reccs}, showing recommendations for summing galaxy alignments along the LOS. $\sigma$ is the standard deviation of $s_\|$-based Gaussian weights for logarithmic bins in $r_p$. When a traditional $\Pi_{\rm max}$ is preferred, the last column provides optimal values in these bins.}
\label{tab:rp_wid1}
\end{table}

\vspace{.1in}\section{Comparing Estimators}
\label{sec:abacus_tests}

\subsection{Halo catalog}
We evaluate the above estimators using halos from the A\textsc{bacus}S\textsc{ummit} N-body simulations \citep{maksimova_abacussummit_2021, garrison_abacus_2021}. These large N body simulations allow us to test many iterations of survey-like samples and precisely compare SNR across methods. We do not expect these simulations to accurately model the alignment of galaxies, which have weaker tidal alignment than halos and depend upon baryon dynamics, even at large scales. However, the relative performance of the estimators is independent of the true IA amplitudes, assuming they have similar scale dependence. We compare different treatments of the LOS direction in narrow $r_p$ bins and the $s_\|$ dependence of the signal in each bin is dominated by shape projections and RSD. 

We create simple mock catalogs from the 25 A\textsc{bacus}S\textsc{ummit} base simulations, generated at $z=0.8$. Each contains 6912$^3$ particles in a 2 $h^{-1}$ Gpc cubic box, which have been used to identify halos with the CompaSO Halo Finder \citep{hadzhiyska_compaso_2021}. We select the largest ones, containing an average of about 10,000 particles per halo. Each of the 25 simulations contains 2.5 million halos, with a comoving density of 3.3 $\times 10^{-4}h^{3}$Mpc$^{-3}$. This is designed to approximately match the density in DESI's Luminous Red Galaxy sample. We create a light cone by placing an observer 2000 $h^{-1}$Mpc away from the center of each box along one axis (corresponding to the comoving distance at z=0.8), and determine a redshift without RSD and one with RSD. 

\begin{figure*}
\centering
\includegraphics[width=.98\textwidth]{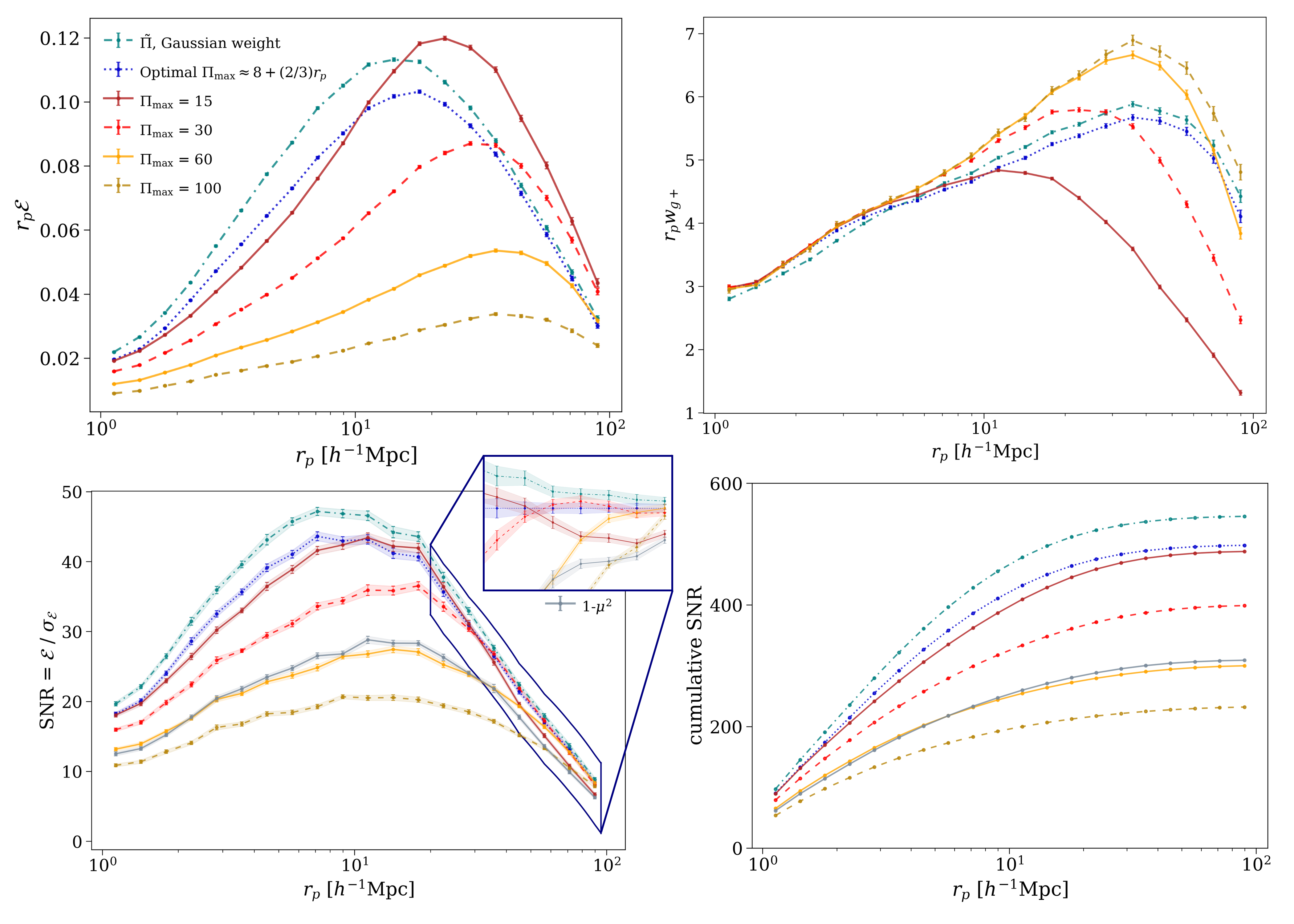}
\caption[]{Comparison between different LOS treatments for $\mathcal{E}_+$ and $w_{g+}$, estimated as $\big(2\Pi_{\rm max} + w_p(r_p)\big)\mathcal{E}_+(r_p)$. The measurements made with various $\Pi_{\rm max}$ choices and averaged over 25 A\textsc{bacus}S\textsc{ummit} simulations are shown in the top two panels. Since the errors are dominated by shape noise, the SNR is the same for the two estimators. The bottom left panel shows the SNR of each measurement, and the bottom right shows the cumulative SNR as $r_p$ increases. The SNR was estimated independently in each simulation and then averaged for this plot. As expected, the SNR is highest for the Gaussian weights across all scales, followed by the optimal $\Pi_{\rm max}$. Among the single $\Pi_{\rm max}$ choices, generally smaller $\Pi_{\rm max}$ perform better, at least at scales bellow 30 $h^{-1}$ Mpc. The insert shows a closer look at where the estimators' SNR cross, although many become statistically indistinguishable at the largest scales.}
\label{fig:snr_comparison}
\end{figure*}

\begin{figure*}
\centering
\includegraphics[width=.9\textwidth]{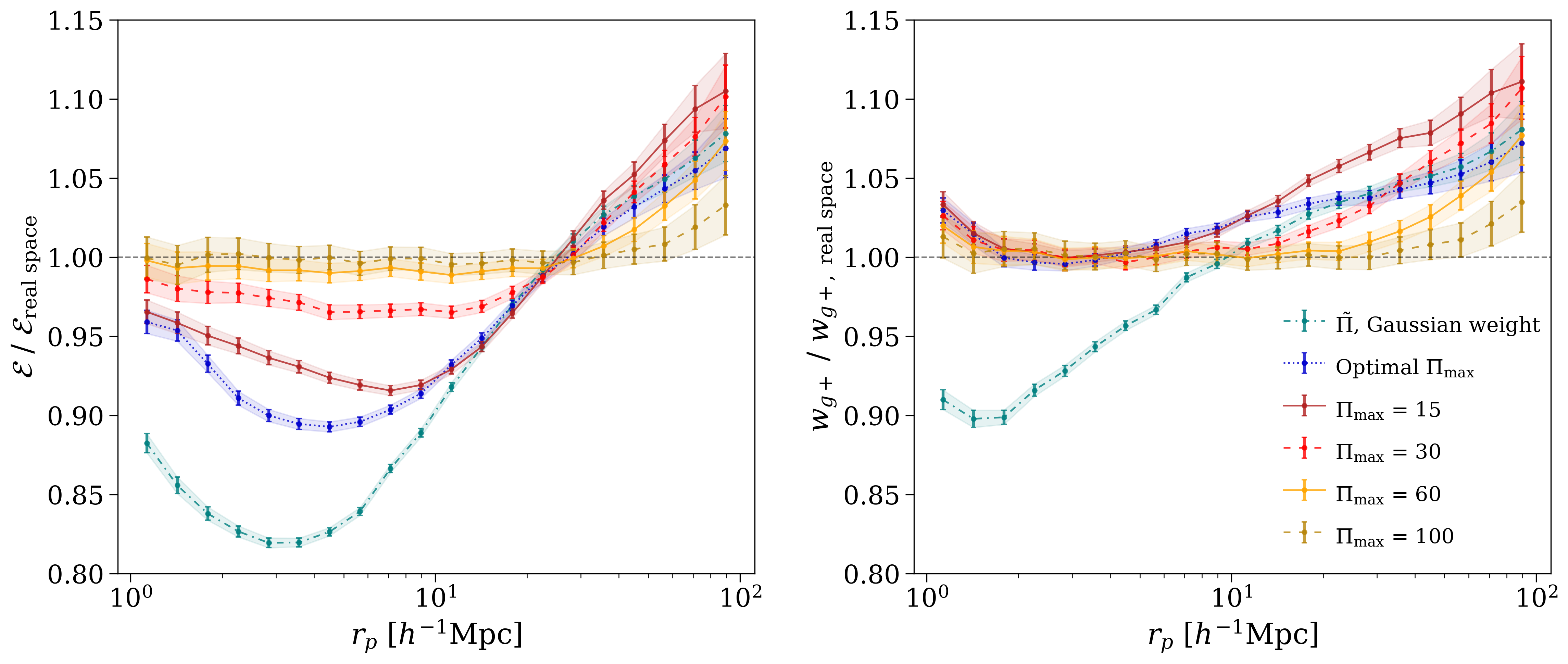}
\caption[]{These plots demonstrate the effect of RSD on each estimator by showing the ratio of its signal in redshift space to real space. In the first plot, $\mathcal{E}$ is lowered on small scales as FOG smears close pairs and their strong correlations along the LOS. This is counteracted by the Kaiser effect which concentrates these correlations along the LOS on larger $r_p$ scales. While $\mathcal{E}$ describes the alignment averaged over each shape-tracer pair, $w_{g+}$ includes 2-point clustering, which counteracts the shift of alignment strength between real and redshift space. As the Gaussian-weighted $\tilde{\Pi}$ includes the most LOS information, it is most sensitive to RSD. While all of these estimators are less sensitive to RSD than 3D correlation functions, models should still include RSD for high-accuracy measurements.}
\label{fig:rsd_comp}
\end{figure*}

\subsection{Alignment Measurements}
To produce the weights displayed in Figures \ref{fig:weight_derivations} - \ref{fig:weight_functions}, we find pairs of galaxies up to projected separations of $r_p< 100 h^{-1}$ Mpc and transverse separations of $|r_\| |<100 h^{-1}$ Mpc in real comoving space. For each pair, we calculate its separation in the plane transverse to the LOS $r_p$ and along the LOS in both real and redshift space, $r_\|$ and $s_\|$. The LOS is defined for each pair as the vector between the observer and pair center. We also obtain the projected ellipticity of each halo along its unique LOS, following \citep{lamman_intrinsic_2023}. 

For comparing alignment estimators, we trim the A\textsc{bacus} catalogs to a uniform survey geometry using RA and DEC within $\pm 10$ deg and redshifts $0.7<z<1.3$. Within these mock catalogs, we measure $\mathcal{E}_+(r_p)$ in each simulation using the $\tilde{\Pi}$ weights described in Section \ref{sec:weighted_pimax}, the variable $\Pi_{\rm max}$ described in Section \ref{sec:optimal_pimax}, and a selection of flat $\Pi_{\rm max}$ cuts. The shapes from the trimmed catalogs are measured relative to the full ones. The error is estimated in each simulation by measuring the signal independently in 100 regions and taking the standard error. This results in 25 determinations of SNR for each estimator. We average and take the standard error of these 25 SNRs to evaluate the performance of each estimator in each $r_p$ bin. The results are unchanged when using bootstrap error estimation from these regions.

Code for these measurements and the associated modeling is publicly available in the repository \href{https://github.com/cmlamman/spec-IA}{github.com/cmlamman/spec-IA}.

\subsection{Results}

The measurements of $\mathcal{E}_+$ made with each $\Pi_{\rm max}$ choice, the corresponding plot for $w_{g+}$, and their average SNR are shown in Figure \ref{fig:snr_comparison}. The LOS weighting $\tilde{\Pi}$ has a 2.3 times improvement over a flat $\Pi_{\rm max}=100h^{-1}$ Mpc, followed closely by using the optimal $\Pi_{\rm max}$ in each $r_p$ bin. As expected, smaller LOS distances tend to be more advantageous at small $r_\|$. However, it's worth noting that $\Pi_{\rm max}=15h^{-1}$ Mpc performs well too, significantly better than the common choice of $\Pi_{\rm max}=100h^{-1}$ Mpc. Since tidal alignment concentrates most of the signal along the $r_p$ axis, a smaller $\Pi_{\rm max}$ is preferred compared to when measuring standard 2-point clustering. Above $r_p=12h^{-1}$Mpc, several estimators perform similarly as the signal has less variation in $r_\|$ at very large $r_p$. All SNR curves on this plot peak around $7-11h^{-1}$Mpc, a scale which depends on our choice of logarithmic binning. The lower panels of Figure \ref{fig:snr_comparison} additionally include a comparison to a simple $1-\mu^2$ weighting, which has a worse SNR than the Gaussian weights because it does not take into account the $r_\|$ dependence of the IA signal within each $r_p$ bin.

To explore the impact of RSD on each of these estimators, we compare their signals in redshift space to real space (Figure \ref{fig:rsd_comp}). Although they are less sensitive to RSD than full 3D correlation functions, all are affected by RSD, which should be included in their modeling for high-accuracy measurements. On small scales, the alignment signal is diluted along the LOS due to the FOG effect. This is counteracted by the change in 2-point clustering, which is enhanced by RSD \citep{nock_effect_2010}. Therefore, $w_{g+}$ is moderately unaffected between $2<r_p<6$ $h^{-1}$Mpc for all flat $\Pi_{\rm max}$, including the optimal (varying) $\Pi_{\rm max}$. The Gaussian-weighted $\tilde{\Pi}$ includes the most LOS information and is most sensitive to RSD, but not to the extent of 3D correlation functions at small scales \citep{singh_increasing_2024}.

\vspace{.1in}\section{Modeling Variable $\mathbf{\Pi_{\rm max}}$}
\label{sec:modeling}
As seen in the top two panels of Figure \ref{fig:snr_comparison}, the modified $\Pi_{\rm max}$ estimators have a different scale dependence than the classic flat $\Pi_{\rm max}$ cuts. This section outlines how to adapt standard intrinsic alignment models to account for these modifications. For the optimal $\Pi_{\rm max}$, modeling follows the same procedure as traditional projected IA statistics, but applied separately to individual $r_p$ bins with their respective $\Pi_{\rm max}$ values. For the $\tilde{\Pi}$ approach with Gaussian weights, the top hat function used for a flat $\Pi_{\rm max}$ cut which selects matter along the LOS must be replaced with a Gaussian. Here we present a modified modeling framework of the most common IA model: nonlinear alignment (NLA), in which the shapes of galaxies are assumed to linearly correlate with the nonlinear matter power spectrum. We don't include RSD in this basic demonstration, but this general approach can be applied to other modeling frameworks.

\begin{figure}
\begin{center}
\includegraphics[width=.48\textwidth]{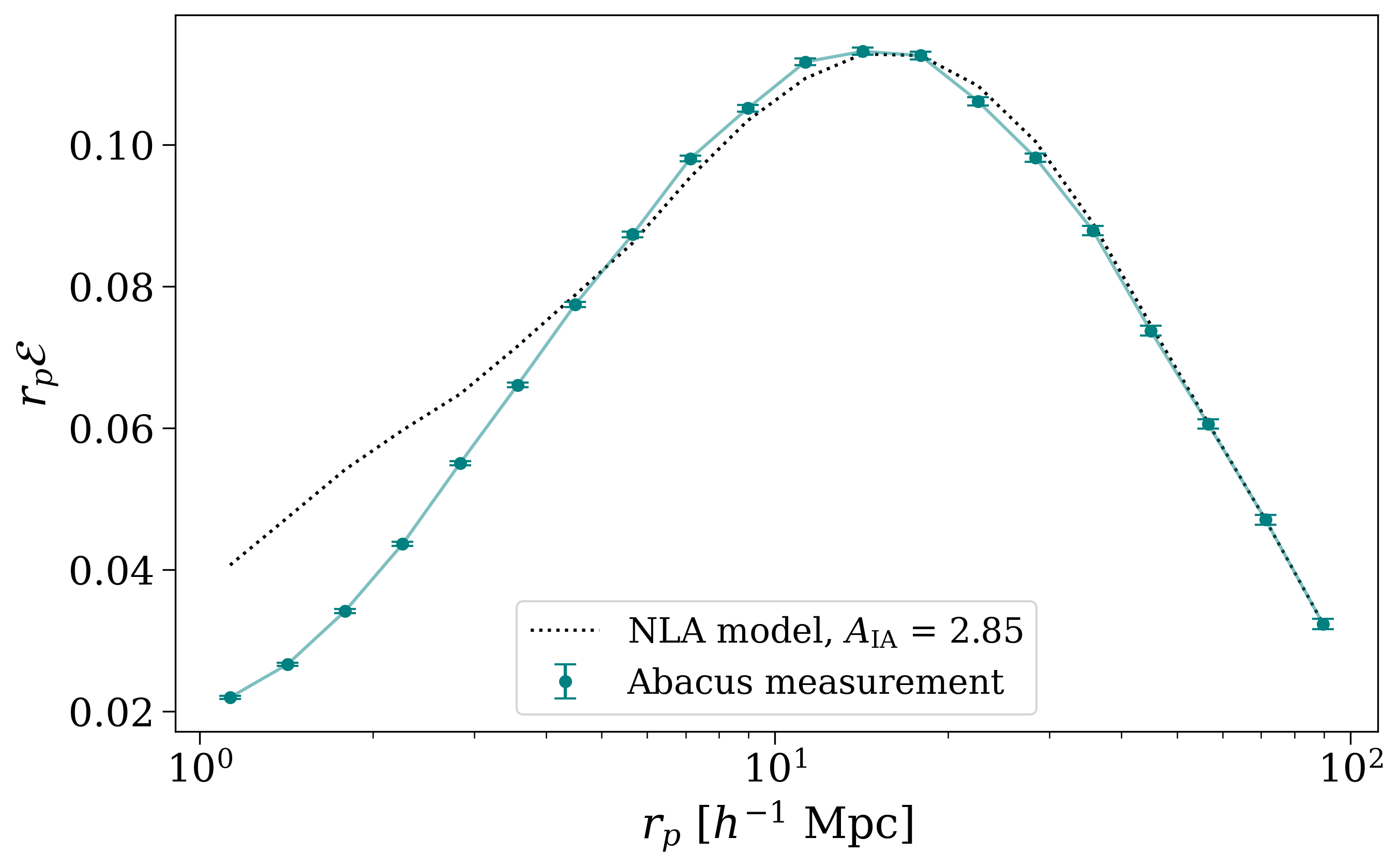}
\end{center}
\caption[]{A comparison of the Gaussian-weighted IA signal shown in Figure \ref{fig:snr_comparison} to the nonlinear alignment model, predicted using the nonlinear A\textsc{bacus} matter power spectrum. The alignment amplitude was estimated from the signal strength above 12 $h^{-1}$Mpc. As expected, this model breaks down at scales below 6 $h^{-1}$Mpc.}
\label{fig:modeling_comp}
\end{figure}

The strength of intrinsic alignments is quantified by the alignment amplitude $A_{\rm IA}$, which describes the response of galaxy shapes to the large-scale gravitational tidal field. The tidal tensor $T_{ij}$ is defined as:

\begin{equation}\label{eq:tidal_tensor}
    T_{ij} = \partial_i \partial_j \phi - \frac{1}{3} \delta^K_{ij} \nabla^2 \phi
\end{equation}
\noindent where $\phi$ is the gravitational potential and $\delta^K_{ij}$ is the Kronecker delta. In Fourier space, this becomes
\begin{equation}
    T_{ij}(\vec r) = \int \frac{d^3k}{(2\pi)^3} \left( \frac{k_i k_j - \frac{1}{3}\delta^K_{ij}k^2}{k^2}\right) \tilde\delta_m(\vec k) e^{i\vec k\cdot\vec r},
\end{equation}
\noindent following \cite{lamman_intrinsic_2023}. Here, $\tilde\delta_m(\vec k)$ is the Fourier transform of the matter density field. In the linear and nonlinear alignment model, galaxy ellipticities are assumed to scale with the tidal field according to:
\begin{equation}\label{eq:E_complex}
    \epsilon = \tau [T_{xx} - T_{yy} + 2iT_{xy}] \equiv \tau \mathbf{T}_\mathcal{E},
\end{equation}
\noindent where $\tau$ is a proportionality constant. The intrinsic shear is related to the alignment amplitude by:

\begin{equation}
    \gamma^I_{ij} = -A_{\rm IA}(z)C_1\frac{\rho_{\text{m},0}}{D(z)}T_{ij}.
\end{equation}
\noindent Here, $\rho_{\text{m},0}$ is the present-day matter density, $D(z)$ is the linear growth factor (normalized so $\bar{D}(z) = (1+z)D(z)$ is unity at present day), and $C_1 = 5\times 10^{-14} M_\odot^{-1}h^{-2}\text{Mpc}^3$ is a historical normalization constant introduced by \cite{brown_measurement_2002}. The relationship between our alignment parameter $\tau$ and the conventional $A_{\rm IA}$ is:
\begin{equation}
    A_\mathrm{IA} (z) = -\frac{\tau}{C_1} \frac{D(z)}{\rho_{\rm m,0}}
\end{equation}

Following \cite{lamman_detection_2024}, we can express the alignment signal in terms of the relevant matter power spectrum:
\begin{equation}
\tilde{\mathcal{E}} = \frac{-\tau}{(2\Pi_{\rm max} + \bar{w}_p)} \int KdK\mathcal{J}_2(K, R) \mathcal{P}_\Pi (K)
\end{equation}

\noindent $\bar{w}_p$ represents the projected 2-point cross-correlation function between the shape and tracer catalogs within an annulus defined by $R_{\rm min}$ and $R_{\rm max}$. The variable $K$ corresponds to the Fourier-space wavenumber in the transverse plane (encompassing $k_x$ and $k_y$), while $k_z$ represents the wavenumber along the LOS direction, with the full 3D wavenumber being $k^2 = K^2 + k_z^2$. $\mathcal{J}_2$ is a modified Bessel function integrated over a given radial bin (see Eq.~\ref{eq:J2}), and $\mathcal{P}_\Pi(K)$ encapsulates the relevant galaxy-matter power spectrum contribution for a given $\Pi_{\rm max}$. 
The expectation value of the cross-correlation between projected shapes and the underlying matter field, denoted as $Q$, is:
\begin{equation}
\tilde{\mathcal{E}}_{\rm model} = \frac{1}{2}\langle \epsilon^* Q + \epsilon Q^* \rangle
\end{equation}
\begin{equation}
Q(R_{\rm bin}, \pm\tilde{\Pi}) = \frac{\int d^3r W(\bar{r})\delta_g e^{2i\theta_r}}{\int d^3r W(\bar{r})(1 + \xi_{\epsilon g})}
\end{equation}
\noindent $Q$ describes the 3D matter field in a specific bin of transverse separation $R_{\rm bin}$ and weighted LOS separation $\tilde{\Pi}(r_\|)$. The function $W(\bar{r})$ serves as a window function that selects this region in configuration space. The LOS weights for each $r_p$ bin are modeled as a normalized Gaussian with standard deviation $\sigma$:
\begin{equation}
    \tilde{\Pi}(s_\|) = \frac{1}{\sigma\sqrt{2\pi}}e^{-(s_\|^2 / 2\sigma^2)}
\end{equation}

\noindent This modified window function now needs to be carried through the expansion of Equation 16:

\begin{equation}\label{eq:E_model_0}
    \begin{split}
    \mathcal{E}_{\rm model} = \frac{-\tau}{\int d^3r W(\bar{r})(1 + \xi_{\epsilon g})}
    \int d^3r W(\bar{r}) \\
    \int \frac{dk_z}{\text{2}\pi} \int KdK J_2(KR)\frac{K^2}{k^2}P_{gm}(k)e^{ik_z z}.
    \end{split}
\end{equation}
The volume element in our weighted space becomes:
\begin{equation}
    \int d^3r W(\bar{r}) = \pi(R_{\rm max}^2 - R_{\rm min}^2)\int r_\|d_\|\tilde{\Pi}(r_\|) % = \int_{R_{\rm min}}^{R_{\rm max}} \int_0^{2\pi}  \int_{-\infty}^{\infty} r_\| r_p \tilde{\Pi}(r_\|) dr_\| dr_p d\theta
\end{equation}
Similarly, the normalization factor accounting for the clustering signal is:
\begin{equation}\int d^3r W(\bar{r})(1+\xi_{\epsilon g})= \pi(R_{\rm max}^2 - R_{\rm min}^2) \int r_\|dr_\|\tilde{\Pi}(r_\|)\big(1 + \xi(r_\|)\big).
\end{equation}
\noindent The radial kernel $\mathcal{J}_2$ remains unchanged from the standard case with a fixed $\Pi_{\rm max}$:
\begin{equation}
\label{eq:J2}
    \mathcal{J}_2(K) = \frac{2}{(R_{\rm max}^2 - R_{\rm min}^2)} \int_{R_{\rm min}}^{R_{\rm max}} RdR J_2(KR),
\end{equation}

 While a top-hat $\Pi_{\rm max}$ cut produces a sinc function in Fourier space, our Gaussian weighting transforms to:

$$\int_{-\infty}^{\infty} dz \tilde{\Pi}(z) \exp^{ik_z z} = \exp^{-k_z^2 \sigma_i^2 / 2}$$.

%http://www.cse.yorku.ca/~kosta/CompVis_Notes/fourier_transform_Gaussian.pdf
\noindent This leads to a modified power spectrum contribution:

\begin{equation}\label{eq:ps_component}
\mathcal{P}_{\tilde{\Pi}}(K) = \frac{1}{{2\pi}}\int dk_z \frac{K^2}{K^2 + k_z^2} P_{gm}\bigg(\sqrt{K^2 + k_z^2}\bigg) \sigma \exp^{-k_z^2 \sigma^2 / 2}.
\end{equation}

\noindent $P_{gm}$ is the galaxy-matter power spectrum, which we estimate using the A\textsc{bacus} matter power spectrum and a galaxy bias of $b=2.7$. For our weighted estimator, the ``effective $\Pi_{\rm max}$'', $(2\Pi_{\rm max} + \bar{w}_p)$, becomes:
\begin{equation}
\tilde{\Pi}_{\rm eff} = 
\frac{\int s_\|dr_\|\tilde{\Pi}_{r_p}(s_\|)\big(1 + \xi(r_p, s_\|)\big)}{\int dr_\|\tilde{\Pi}_{r_p}(s_\|)}
\end{equation}
Our final expression for the alignment signal is therefore:
\begin{equation}
\mathcal{E}(r_p) = \frac{-\tau}{\tilde{\Pi}_{\rm eff}} \int KdK\mathcal{J}_2(K, r_p) \mathcal{P}_{\tilde{\Pi}} (K).
\end{equation}

Figure \ref{fig:modeling_comp} compares the LOS-weighted $\mathcal{E}_+$ to the normalized model, which reproduces the signal above $r_p=6h^{-1}$Mpc.

IA models often use the Limber approximation, assuming that only the component of the tidal field transverse to the line of sight, $k_z=0$ modes, will contribute \citep{blazek_tidal_2015}. In this case, Equation \ref{eq:ps_component} simply becomes the galaxy power spectrum, $\mathcal{P}_{\tilde{\Pi}}(K) = P_{gm}(K)$. This is valid when $r_p \ll \Pi_{\rm max}$, and can be sufficient for $\Pi_{\rm max}= 30 - 100 h^{-1}$Mpc alignment signals at small $r_p$. The $\Pi_{\rm max}$ recommendations we provide here are smaller and should not be modeled using the Limber approximation. While modeling the full power spectrum is computationally more expensive, our LOS weighting comes with its own computational advantage. The Gaussian weights eliminate the oscillatory behavior of the sinc function that arises from a sharp $\Pi_{\rm max}$ cut, leading to more numerically stable and efficient integration. 

%\needspace{5\baselineskip}
\vspace{.1in}\section{Conclusion}

The advent of large spectroscopic surveys necessitates a reevaluation of how to efficiently handle LOS information when measuring intrinsic alignments. Although 3D estimators can achieve higher SNR than projected correlations, they come with increased sensitivity to LOS effects, more complex modeling requirements, and are more difficult to directly relate to shear measurements. We present an alternative set of projected correlation functions that better capture LOS information. These result in similar SNR improvements compared to classic estimators as Singh et al.'s weighted quadrupole, but with the benefits of a projected statistic.

Our main results demonstrate that, at minimum, direct IA detections should adopt smaller $\Pi_{\rm max}$ values than the conventional 60-100 $h^{-1}$Mpc. A variable $\Pi_{\rm max}$ that scales with projected separation provides further improvement, approximately following $\Pi_{\rm max}= 3 + (2/5)r_p$ $h^{-1}$ Mpc for $r_p > 2h^{-1}$ Mpc and increasing at smaller scales to better average over FOG effects. SNR can be further optimized by utilizing our LOS weighting scheme, $\tilde{\Pi}$, which shows a 2.3 times improvement in SNR over flat $\Pi_{\rm max}=100h^{-1}$Mpc cuts. Although the most optimal widths of Gaussian weights at very small scales may be sample-dependent, our approach to derive them is general and can be and applied at any scale. The performance differences between these approaches become marginal over projected separations of $12 h^{-1}$Mpc. While modeling remains challenging at small separations, considerable information resides there. Our modified $\Pi_{\rm max}$ approach can help extract IA correlations in the small scales where modeling RSD becomes impractical for full 3D correlations. As the Kaiser effect only impacts IA beyond linear order, RSD alone may not be a compelling reason to adopt these estimators at large scales. Situations where they may be advantages over 3D estimators at large scales include: large redshift uncertainties, exploring nonlinear correlations, ease of modeling, and translating to projected cosmic shear statistics.

It is worth emphasizing that our recommendations for optimal LOS treatment are effectively independent of features within the IA signal itself. Although, the specific values of the Gaussian widths should increase with satellite fraction. An ideal, but perhaps ineffectual, approach would be to measure IA with basic weights and then re-iterate with new weights based on these measurements. However, the choices for optimal $\Pi_{\rm max}$ and weights are dominated by the projection of shapes and RSD. They are also independent of various convention choices, including: correlation functions (such as $\mathcal{E}$ vs $w_{g+}$), shape definitions, and object type (galaxies vs halos vs ensembles of galaxies). Although we focus on shape-density correlations in this work, similar principles apply to shape-shape correlations such as $w_{++}$, with the same RSD model combined with a $(1-\mu^2)^2$ dependence.

Maximizing the precision of IA measurements directly improves constraints on the alignment amplitude. By more efficiently handling LOS information, these estimators will enhance the cosmological information extracted from large spectroscopic surveys, benefiting both weak lensing analyses and the exploration of direct cosmological applications of IA.

\section*{Acknowledgements}

CL wishes to acknowledge useful conversations with Teppei Okumura and Chris Hirata.

This material is based upon work supported by the U.S.\ Department of Energy under grants DE-SC0013718 and DE-SC0024787, NASA under ROSES grant 12-EUCLID12-0004 and the Roman Research and Support Participation program grant 80NSSC24K0088, the U.S.\ National Science Foundation under award AST-2206563, and the Simons Foundation. CL is additionally supported by an NSF Postdoctoral Fellowship under award 2502789.

%%%%%%%%%%%%%%%%%%%%%%%%%%%%%%%%%%%%%%%%%%%%%%%%%%
\section*{Data Availability}
A\textsc{bacus}S\textsc{ummit} data products are publicly available at \href{https://abacusnbody.org/}{abacusnbody.org} \citep{maksimova_abacussummit_2021}.

Code used in this work is publicly available in the repository \href{https://github.com/cmlamman/spec-IA}{github.com/cmlamman/spec-IA}.

%%%%%%%%%%%%%%%%%%%% REFERENCES %%%%%%%%%%%%%%%%%%
\bibliographystyle{mnras}
\bibliography{references} % if your bibtex file is called example.bib

\end{document}